\documentclass[aps,prb,reprint,a4paper,floatfix,superscriptaddress,amsmath,amssymb,amsfonts]{revtex4-1}
\usepackage{newtxtext,newtxmath}
\usepackage[utf8]{inputenx} 
\usepackage{graphicx}
\usepackage{tabularx, booktabs}
\usepackage{xspace}
\usepackage[dvipsnames]{xcolor}
\usepackage[pdftex]{hyperref}
\usepackage{verbatim}
\hypersetup{
	pdftitle={Band gap widening and behavior of Raman-active phonon modes of cubic single-crystalline (In,Ga)$_2$O$_3$ alloy films},
	colorlinks,
	citecolor=blue
	}
\usepackage[
,textwidth=17.5cm
,textheight=23.7cm
,verbose
,dvips
]{geometry}

\begin{document}

\title{Band gap widening and behavior of Raman-active phonon modes of cubic single-crystalline (In,Ga)$_2$O$_3$ alloy films}

\author{Johannes Feldl}
%\thanks{These two authors contributed equally}
\affiliation{Paul-Drude-Institut f\"ur Festk\"orperelektronik, Leibniz-Institut im Forschungsverbund Berlin e.~V., Hausvogteiplatz 5--7, 10117 Berlin, Germany}
\author{Martin Feneberg}
%\thanks{These two authors contributed equally}
\affiliation{Institut f\"ur Physik, Otto-von-Guericke-Universit\"at Magdeburg, Universit\"atsplatz 2, 39106 Magdeburg, Germany}
\author{Alexandra Papadogianni}
\affiliation{Paul-Drude-Institut f\"ur Festk\"orperelektronik, Leibniz-Institut im Forschungsverbund Berlin e.~V., Hausvogteiplatz 5--7, 10117 Berlin, Germany}
%\author{Shigenori Ueda}
%\affiliation{National Institute for Materials Science, Tukuba, Ibaraki 305-0044, Japan}
\author{Jonas L\"ahnemann}
\affiliation{Paul-Drude-Institut f\"ur Festk\"orperelektronik, Leibniz-Institut im Forschungsverbund Berlin e.~V., Hausvogteiplatz 5--7, 10117 Berlin, Germany}
\author{Takahiro Nagata}
\affiliation{National Institute for Materials Science, Tsukuba, Ibaraki 305-0044, Japan}
\author{Oliver Bierwagen}
\affiliation{Paul-Drude-Institut f\"ur Festk\"orperelektronik, Leibniz-Institut im Forschungsverbund Berlin e.~V., Hausvogteiplatz 5--7, 10117 Berlin, Germany}
\author{R\"udiger Goldhahn}
\affiliation{Institut f\"ur Physik, Otto-von-Guericke-Universit\"at Magdeburg, Universit\"atsplatz 2, 39106 Magdeburg, Germany}
\author{Manfred Ramsteiner}
\email{ramsteiner@pdi-berlin.de}
\affiliation{Paul-Drude-Institut f\"ur Festk\"orperelektronik, Leibniz-Institut im Forschungsverbund Berlin e.~V., Hausvogteiplatz 5--7, 10117 Berlin, Germany}

\begin{abstract} 
The influence of Ga incorporation into cubic In$_2$O$_3$ on the electronic and vibrational properties is discussed for (In$_{1-x}$,Ga$_x$)$_2$O$_3$ alloy films grown by molecular beam epitaxy. Using spectroscopic ellipsometry, a linear dependence of the absorption onset on the Ga content $x$ is found with a blueshift of up to 150 meV for $x = 0.1$. Consistently, the fundamental band gap exhibits a blueshift as determined by hard X-ray photoelectron spectroscopy. The dependence of the absorption onset and the effective electron mass on the electron concentration is derived from the infrared dielectric functions for a Sn doped alloy film. The influence of alloying on phonon modes is analyzed on the basis of Raman spectroscopic measurements. The frequencies of several phonon modes are identified as sensitive measures for the spectroscopic determination of the Ga content.
\end{abstract}

\maketitle

The sesquioxides In$_2$O$_3$ and Ga$_2$O$_3$ are wide band gap semiconductors which hold promise for applications in the field of transparent electronics.
\citep{bierwagen_2015,galazka_2018,vonwenckstern_2017} Since the fundamental band gaps of these two binary compounds in their most stable structural phases span the range from 2.7~eV (In$_2$O$_3$) to 4.8~eV (Ga$_2$O$_3$), an extended tunability of the electronic properties is expected for (In,Ga)$_2$O$_3$ alloys.\citep{hill_1974} However, the crystal structure of In$_2$O$_3$ is cubic bixbyite whereas Ga$_2$O$_3$ assumes a monoclinic structure ($\beta$-Ga$_2$O$_3$) in thermal equilibrium.\citep{peelaers_2015} Accordingly, the knowledge about the miscibility of the binary compounds for a specific crystal structure is of crucial importance.\citep{maccioni_2015,maccioni_2016,maccioni_2016a,wouters_2020} Furthermore, material properties such as the electronic band structure and the behavior of phonon modes in (In,Ga)$_2$O$_3$ alloys are of fundamental interest. In particular for monoclinic alloys with Ga contents above 60\%, experimental results regarding physical properties like the composition-dependent electronic band gap and phonon frequencies can be found in the literature for films grown by various methods.\citep{oshima_2008,yang_2009,kokubun_2010,baldini_2014,zhang_2014,nagata_2020,vigreux_2001,kranert_2014} 
However, for cubic (In,Ga)$_2$O$_3$ with Ga contents below 30\%, reports on phonon modes\citep{kranert_2014,regoutz_2015} are still scarce and the literature on the dependence of the optical band gap on the alloy composition does not draw a consistent picture. There are experimental studies of (In,Ga)$_2$O$_3$ alloys where a band gap narrowing for low Ga contents were observed,\citep{regoutz_2015,minami_1996,chun_2004,kudo_1998} against the expected band gap widening. Density functional theory calculations for the optically allowed band gap of (In,Ga)$_2$O$_3$ alloys for low Ga amounts, predict both band gap widening\citep{regoutz_2015,maccioni_2016a} and narrowing.\citep{peelaers_2015} In addition, other experimental studies could show an increase in optical absorption onset, with increasing Ga content, though the step size in varying the cation ratio Ga/(Ga+In) were rather coarse.\citep{oshima_2008,yang_2009,zhang_2014,yang_2011,swallow_2021} It has to be considered, that studying nanowires\citep{chun_2004} or sintered ceramics\citep{regoutz_2015} make the determination of bulk properties difficult.
In this respect, high-quality single-crystalline films are much better suited, particularly in the regime of low Ga content, which is most controversially discussed. Regarding the desired control of the conductivity in (In,Ga)$_2$O$_3$ alloys, as established for In$_2$O$_3$ by using the n-type dopant Sn,\citep{granqvist_2002,groth_1966} no report can be found in the literature.

In this work, we use spectroscopic ellipsometry, hard X-ray photoelectron spectroscopy (HAXPES) and Raman spectroscopy to study the electronic and vibrational properties of high quality cubic single-crystalline (In,Ga)$_2$O$_3$ synthesized by molecular beam epitaxy (MBE). With increasing Ga contents up to 10\%, monotonic blueshifts of the fundamental band gap, the optical absorption onset as well as the frequencies of selected optical phonon modes are revealed. Utilizing Sb doping, it is furthermore demonstrated that the advantage of band gap engineering by alloying can be transferred to transparent conducting oxides (TCO).

The investigated (In$_{1-x}$Ga$_x$)$_2$O$_3$ films were grown on (111) surfaces of yttria-stabilized zirconia (YSZ) substrates by plasma-assisted molecular beam epitaxy at a growth temperature of 600~°C. After growth, the nominally undoped samples were thermally annealed under an oxygen atmosphere leading to carrier concentrations\citep{bierwagen_2015} of several 10$^{17}$~cm$^{-3}$ as determined by Hall effect measurements (see Tab.~\ref{Samples}). For studying free-carrier related phenomena, a Sn doped alloy film with a nominal Ga content of 8\% was grown. All films are cubic (bixbyite structure) with Ga contents of $0 \leq x \leq 0.1$ according to energy-dispersive X-ray spectroscopy (EDX) with an uncertainty in $x$ of 0.01. More details regarding the growth conditions, the structural properties and the influence of annealing conditions on the carrier concentration can be found in Ref.~\onlinecite{papadogianni_2021}. In addition, a high quality single crystal bulk In$_2$O$_3$, grown from the melt by the so-called levitation-assisted self-seeding crystal growth,\citep{galazka_2014} serving as a reference for binary In$_2$O$_3$ was examined. For the investigation of the optical properties, two different spectroscopic ellipsometers were employed. In the infrared spectral range (from 300~cm$^{-1}$ to 8000~ cm$^{-1}$ with the range below 800~cm$^{-1}$ being not analyzable in a satisfactory manner), a Woollam IR-VASE based on a Fourier transform spectrometer was used, while in the visible and ultraviolet spectral range (from 0.5 to 6.5~eV) a Woollam VASE equipped with autoretarder was employed. The ellipsometric parameters were recorded at three angles of incidence (50°, 60°, and 70°) in order to increase the reliability of the obtained dielectric functions (DFs). The DFs were obtained by multilayer modeling of the ellipsometric data,\citep{goldhahn_2003} where the surface roughness was modeled using the Bruggeman effective medium approximation.\citep{bruggeman_1935} The DF of the layer under study was fitted employing a point-by-point fit, i.~e. without making assumptions about the line shape of the DF. Raman measurements were performed in backscattering geometry from the surfaces of the epitaxial films. The 325-nm (3.81~eV) line of a He-Cd laser was used for optical excitation. The incident laser light was focused by a microscope objective onto the sample surfaces. The backscattered light was collected by the same objective without analysis of its polarization, spectrally dispersed by an 80-cm spectrograph (LabRam HR, Horiba/Jobin Yvon) and detected with a liquid-nitrogen-cooled charge-coupled device (CCD). Low temperature measurements were performed in a continuous-flow cryostat. Polarization dependent Raman experiments were performed with excitation by the 473~nm (2.62~eV) line of a solid-state laser. For the Raman shift calibration, a gas-discharge lamp and a silicon reference sample were used. Peak positions were obtained by fitting with Voigt line-shape functions. HAXPES measurements were performed at room temperature at the revolver undulator beamline at BL15XU of SPring-8 using 5.95~keV hard X-rays.\citep{ueda_2010} Details of the experimental setup of HAXPES at the beamline is described elsewhere.\citep{ueda_2013} A high-resolution hemispherical electron analyzer (VG Scienta R4000) was used to detect the photoelectrons. The total energy resolution of HAXPES was set to 240~meV. The sample was contacted to the system ground to align its Fermi level with that of the spectrometer. To determine the absolute binding energy, the X-ray photoelectron spectroscopy data were calibrated against the Au 4f$_{7/2}$ peak (84.0~eV) and the Fermi level of Au. Peak fitting of the XPES data was carried out using the Voigt function after subtracting the Shirley-type background by KolXPD software.\citep{shirley_1972,kolibrik} Additional information about the fitting procedure can be found in the supplement.

%======================================================================
\begin{table}
\caption{\label{Samples}Nominal Ga content ($x$), film thickness ($t$), carrier concentration ($n$) for undoped (In$_{1-x}$Ga$_x$)$_2$O$_3$ films and, shift of the optical absorption onset $\Delta E_{\mathrm{opt}}$ relative to that of bulk In$_2$O$_3$.}
\begin{ruledtabular}
\begin{tabular}{cccc}
$x$ & $t$ (nm) & $n$ (10$^{17}$ cm$^{-3}$) & $\Delta E_{\mathrm{opt}}$ (meV) \\
\hline
0 & 647  & 6.6 & -\\
0.05 & 667 & 6.1 & 70 \\
0.08 & 703 & 9.5 & 110 \\
0.1 & 638 & 10.8 & 150\\
\end{tabular}
\end{ruledtabular}
\end{table}
%======================================================================

 %%Fig. 1
\begin{figure}%{r}{0.55\textwidth}
%\vspace*{-0.5cm}
\includegraphics*[width=0.48\textwidth]{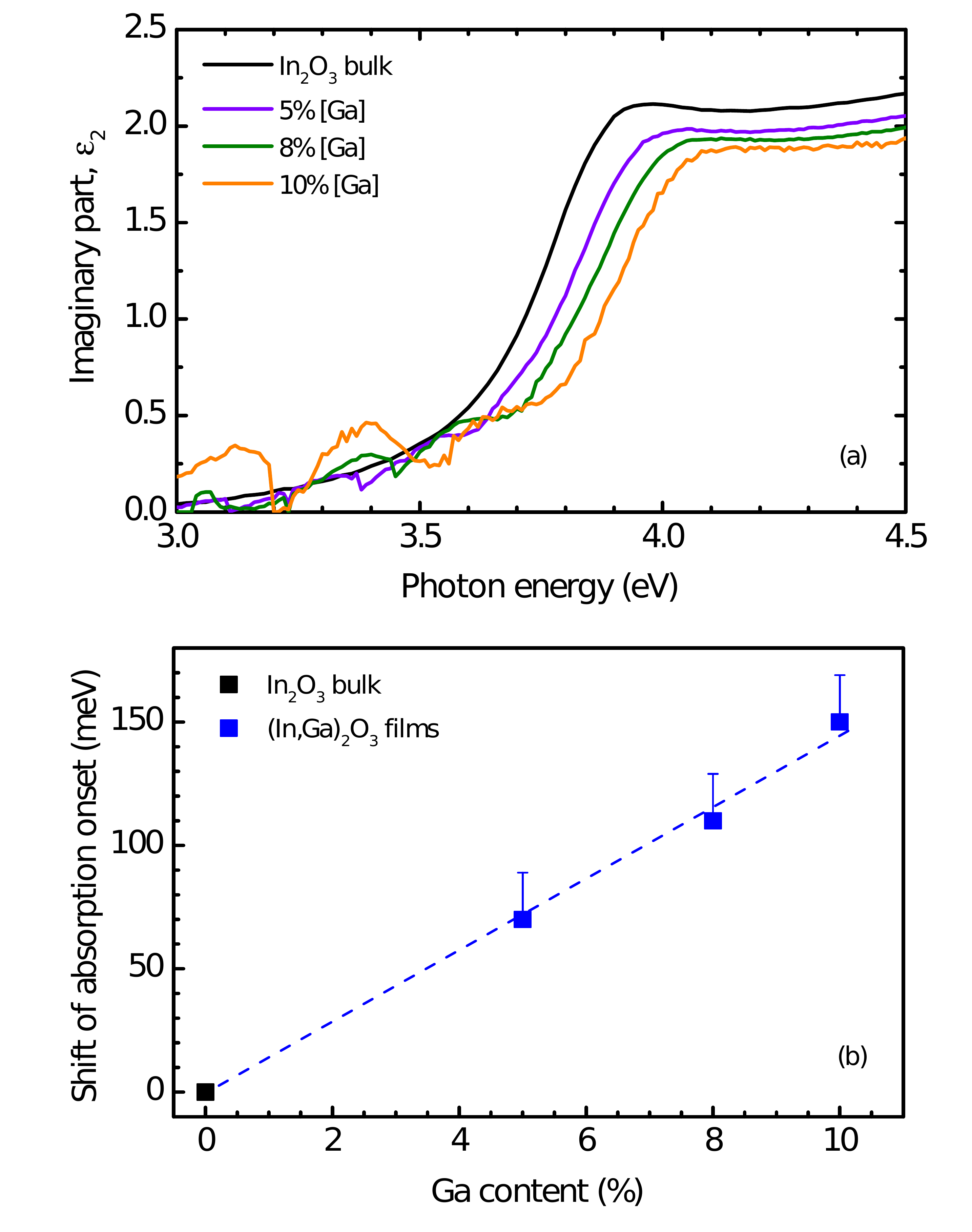}
\caption{(a) Point-by-point fitted imaginary part of the dielectric functions in the spectral range between 3.0 and 4.5~eV for nominally undoped (In,Ga)$_2$O$_3$ films with different Ga contents. (b) Energy shift of the absorption onset with respect to that of cubic In$_2$O$_3$ as a function of Ga content. The error in comparing absorption onsets for different Ga concentration thin films is low since the slope of the imaginary parts of the DFs remain nearly constant. We thus quantify the error margin to be below 20~meV. Error bars indicate an estimated correction by the influence of strain on the absorption onset as described in the text.} 
\label{DF}
%\vspace*{-0.5cm}
\end{figure}
% Fig1 from IGO\impart/Folder1/
%%======================================================================

The influence of the Ga incorporation into In$_2$O$_3$ on the electronic band structure has been investigated by spectroscopic ellipsometry. Figure~\ref{DF}(a) displays the imaginary parts of the dielectric functions for the undoped (In,Ga)$_2$O$_3$ films in the spectral range between 3.0 and 4.5~eV. The binary film exhibits an onset of optical absorption at about 3.8~eV as previously reported for undoped or low doped In$_2$O$_3$.\citep{feneberg_2016} Due to the dipole forbidden nature of the fundamental band gap, this absorption onset is related to the dipole allowed transition from the conduction band minimum to a lower lying valence band.\citep{walsh_2008} The dielectric functions of the alloy films provide clear evidence for a blueshift of the onset of optical absorption with increasing Ga content. The shift of the absorption onset as a function of the Ga content shown in Fig.~\ref{DF}(b) reveals a nearly linear relationship with a slope of $\Delta E/\Delta x =$~1450 meV. Our finding is in strong contrast to the conclusions of Ref.~\onlinecite{regoutz_2015,chun_2004} for polycrystalline films, where even a redshift is claimed as a consequence of the incorporation of Ga into In$_2$O$_3$. The conclusion on bulk properties from those works may be of limited validity due to the influence of grain boundaries in the polycrystalline material\citep{regoutz_2015} and the complex optical path in the nanowire mesh,\citep{chun_2004} respectively. For our oxygen annealed samples with carrier concentrations of no more than 10$^{18}$~cm$^{-3}$ (see Tab.~\ref{Samples}), carrier induced effects such as the Burstein-Moss shift can be neglected.\citep{feneberg_2016}  Regarding the possible influence of residual lattice strain on the optical absorption onset of the alloy films, it has to be considered that for In$_2$O$_3$ films grown under the same conditions, maximum tensile in-plane strain values of about 0.2\% have been determined by X-ray diffraction (XRD) measurements (cf. Ref.~\onlinecite{papadogianni_2021}) for film thicknesses above 200~nm. Assuming such a strain level for the present alloy films, a strain induced redshift of the optical absorption onset by 19~meV would have to be taken into account according to the results of Ref.~\onlinecite{walsh_2011} obtained for In$_2$O$_3$. 
The data in Fig.~\ref{DF}(b) are those values, determined by spectroscopic ellipsometry and corrected of the influence of the corresponding Burstein-Moss shift and band gap renormalization. The error bars show the effect of the estimated strain-induced shift taken into account. In order to do so, the data given in Ref.~\onlinecite{walsh_2011} have been corrected for the influence of free charge carriers according to Ref.~\onlinecite{feneberg_2016}.

%======================================================================
 %%Fig. 2
\begin{figure}%{r}{0.55\textwidth}
%\vspace*{-0.5cm}
\includegraphics*[width=0.48\textwidth]{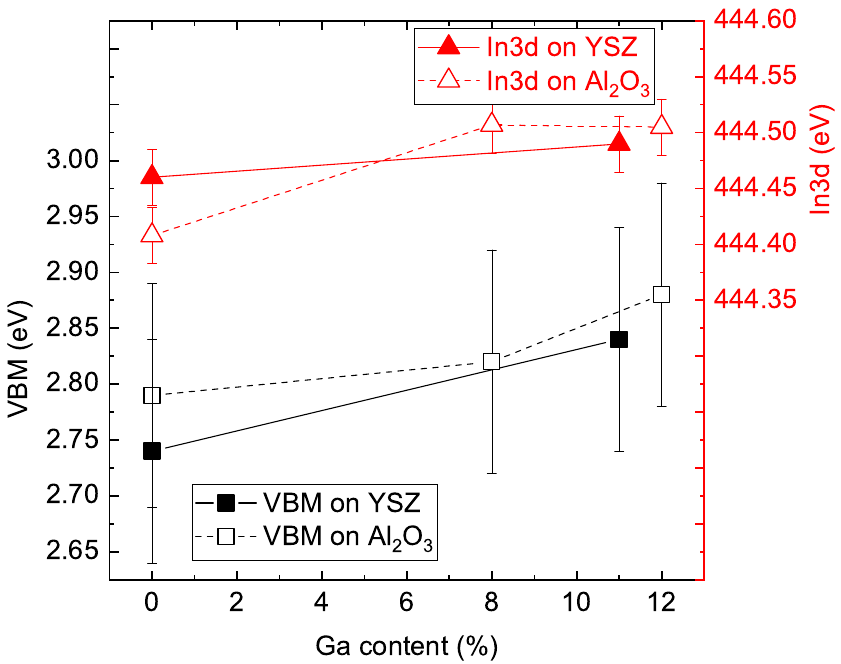}
\caption{HAXPES results for the valence band maximum and In 3d core-level binding energy. The binding energy of zero corresponds to the Fermi level.} 
\label{VBM}
%\vspace*{-0.5cm}
\end{figure}
% Graph8 - Graph5 - Kopie - Kopie from In2O3 and IGO\HAXPES Japan\haxpes_shifts_fundamental_ms/manuscript_gap/
%%========================================

For the investigation of the fundamental band gap of (In,Ga)$_2$O$_3$ alloys, we employed hard X-ray photoelectron spectroscopy. 
Figure~\ref{VBM} shows the valence band maximum (VBM) as well as In 3d core-level binding energy extracted by HAXPES as a function of Ga content for various (In$_{1-x}$Ga$_x$)$_2$O$_3$ films grown on YSZ and $c$--plane Al$_2$O$_3$ substrates grown under conditions similar to those studied by ellipsometry and Raman scattering. The binding energy of zero corresponds to the Fermi level. Based on the electron concentrations measured in our films (see Table~\ref{Samples}) and band parameters taken from In$_2$O$_3$,\citep{feneberg_2016} the Fermi level lies in the vicinity of the conduction band minimum in the bulk. Hence, the VBM binding energy corresponds to the fundamental band gap of the investigated films, within the limits of the measurement accuracy. Despite the large experimental uncertainty, the VBM binding energy is found to increase monotonously with Ga content in a similar fashion as the optical absorption onset (see Fig.~\ref{DF}). In addition, the increasing core-level binding energy largely supports the same trend. Hence, neither the optical nor the fundamental band gap support the initial decrease of the band gap with Ga content reported in Refs.~\onlinecite{regoutz_2015,peelaers_2015,chun_2004}.

%%
%======================================================================
 %%Fig. 3
\begin{figure}%{r}{0.55\textwidth}
%\vspace*{-0.5cm}
\includegraphics*[width=0.48\textwidth]{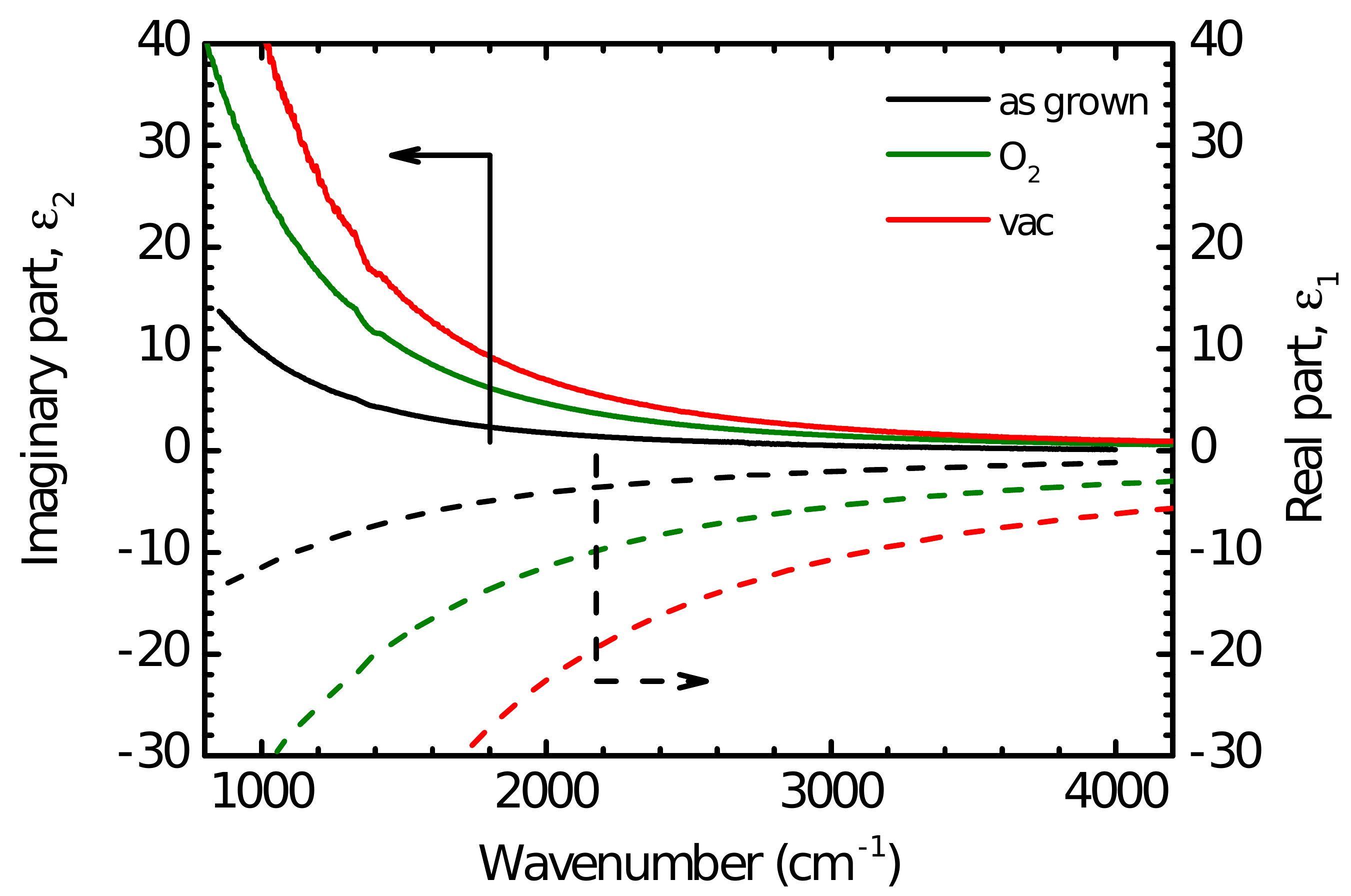}
\caption{Point-by-point fitted infrared dielectric functions (solid lines: real parts, dashed lines: imaginary parts) of a $n$-type doped (In,Ga)$_2$O$_3$:Sn film ($x = 0.08$) before (as-grown) and after oxygen and vacuum annealing.} 
\label{IR}
%\vspace*{-0.5cm}
\end{figure}
% Fig2 from IGO\Fig2wavenumber/Folder1/
%%========================================

%======================================================================
\begin{table*}
\caption{\label{Carriers}Carrier concentrations ($n$) and mobilities ($\mu$) determined by Hall effect measurements, Burstein Moss shift of the onset of optical absorption ($\Delta E_{\text{BM}}$) as well as plasma frequency ($\omega_{\text{P}}$), corresponding damping constant ($\gamma_{\text{P}}$) and effective electron mass ($m_{\text{eff}}$) derived by the analysis of the dielectric function for a (In,Ga)$_2$O$_3$:Sn film with a Ga content of 8\% before and after oxygen or vacuum annealing.}
\centering
\begin{ruledtabular}
\begin{tabular}{ccccccc}
(In,Ga)$_2$O$_3$:Sn & $n$ (10$^{19}$ cm$^{-3}$) & $\Delta E_{\text{BM}}$ (meV) & $\omega_{\text{P}}^2$ (10$^6$ cm$^{-2}$) & $m_{\text{eff}}$ & $\mu$ (cm$^2$/Vs) & $\gamma_{\text{P}}^{-1}$ (10$^{-3}$ cm) \\
\hline
as-grown & 4.4 & 0 & 19.98 & 0.20 & 38.4 & 1.25 \\
oxygen-annealed & 16.5 & 200 & 53.29 & 0.26 & 38.3 & 1.21 \\
vacuum-annealed & 35.0 & 350 & 88.81 & 0.33 & 50.8 & 1.43 \\
\end{tabular}
\end{ruledtabular}
\end{table*}

%======================================================================

%======================================================================
 %%Fig. 4
\begin{figure}%{r}{0.55\textwidth}
%\vspace*{-0.5cm}
\includegraphics*[width=0.48\textwidth]{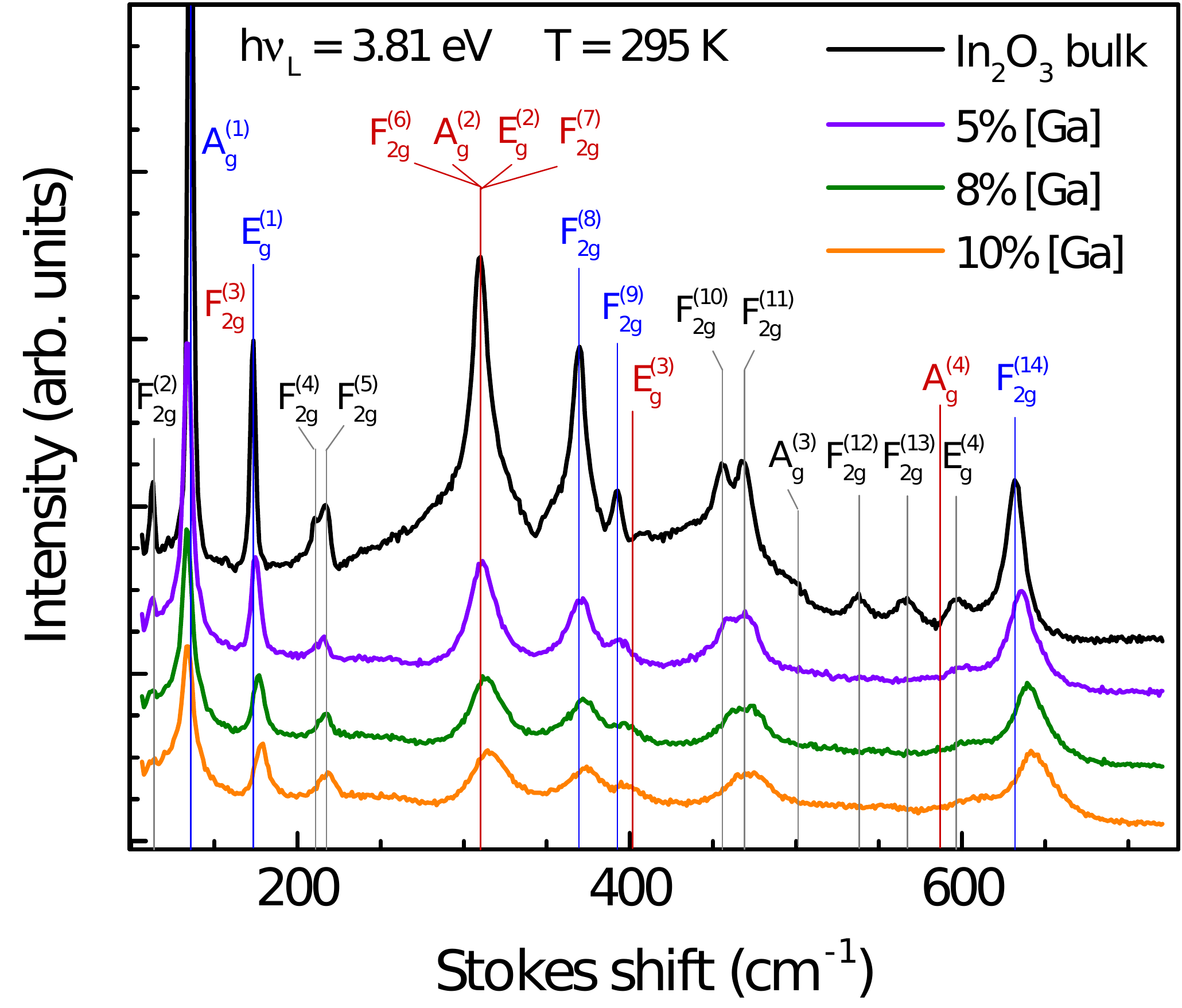}
\caption{Room-temperature (295~K) Raman spectra of (In$_{1-x}$Ga$_x$)$_2$O$_3$ films for different nominal Ga contents $x$ and a reference In$_{2}$O$_3$ bulk sample excited at a photon energy of 3.81~eV. The frequencies of the phonon modes marked by blue vertical lines and labels are displayed in Fig.~\ref{Phonons}. The remaining modes are marked by black color, except those which could not be detected or whose symmetry assignment was not unambiguous.The latter phonon lines are marked in red.} 
\label{Spectra}
%\vspace*{-0.5cm}
\end{figure}
% IGORamanSpectra from \InGa2O3\RT_325nm_Modes/July2020_allSamples_one_cali_RT/Figures
%%======================================================================

A Sn-doped (In$_{0.92}$Ga$_{0.08}$)$_2$O$_3$ was grown to investigate the optical absorption and effective electron mass as a function of electron concentration. The carrier concentration in the Sn doped alloy film could be varied by different annealing conditions (see Tab.~\ref{Carriers}). In this way, we could study the impact of different doping levels at exactly the same Ga content. From the dielectric functions in the spectral range between 3.0 and 4.5~eV (not shown here), a shift of the absorption onset is deduced (see Tab.~\ref{Carriers}) which monotonically increases with the carrier concentration and is quantitatively comparable to the results obtained for cubic In$_2$O$_3$.\citep{feneberg_2016} The parameter-free point-by-point dielectric functions in the infrared spectral range are shown in Fig.~\ref{IR}. Using the Drude model for the free-carrier contribution to line-shape fit the complex DFs, we analyzed the measured data along the lines of Ref.~\onlinecite{feneberg_2016}. The obtained plasma frequencies ($\omega_{\text{P}}$) and the corresponding damping constants ($\gamma_{\text{P}}$) as well as the deduced effective electron masses ($m_{\text{eff}}$) are given in Tab.~\ref{Carriers}. For a better comparison with the Hall effect data, it is taken into account that the carrier concentration (mobility) is proportional to $\omega_{\text{P}}^2$ ($\gamma_{\text{P}}^{-1}$).\citep{feneberg_2016} The plasma frequency is found to monotonically increase with the carrier concentration and the derived effective masses depend on the doping level, in accordance with the case of cubic In$_2$O$_3$.\citep{feneberg_2016,stokey_2021} However, a significant change of the effective masses induced by Ga incorporation could not be deduced. Regarding the damping constant $\gamma_{\text{P}}$, no clear relationship with the carrier mobility could be observed. Further work is necessary to explain this deviation from the expected behavior.  

%======================================================================
 %%Fig. 5
\begin{figure}%{r}{0.55\textwidth}
%\vspace*{-0.5cm}
\includegraphics*[width=0.5\textwidth]{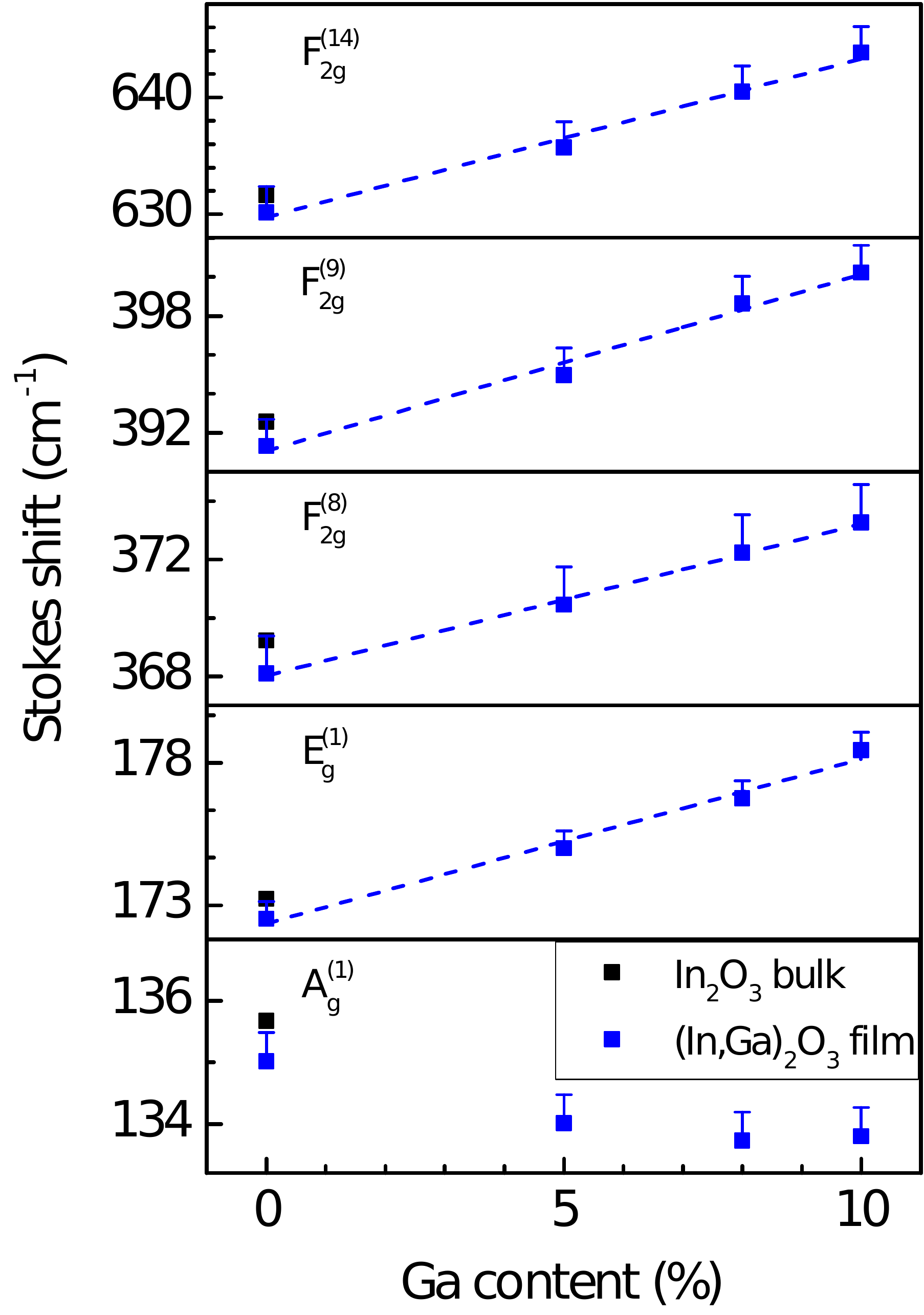}
\caption{Room-temperature Phonon mode frequencies observed in Raman spectra of (In$_{1-x}$Ga$_x$)$_2$O$_3$ films as well as in bulk In$_{2}$O$_3$ as a function of the nominal Ga content ($x$).} 
\label{Phonons}
%\vspace*{-0.5cm}
\end{figure}
% allModesNEWEDXabsolute from \InGa2O3\RT_325nm_Modes/July2020_allSamples_one_cali_RT/corrected_results_merged
%%======================================================================

\begin{table*}
\caption{\label{PhononFrequencies}Frequencies of Raman active phonon modes observed for excitation at 3.81~eV at sample temperatures of 10~K and 295~K. Presented are modes which could be analyzed at least for one alloy film. The uncertainty of the peak position is estimated to be 0.5~cm$^{-1}$, except for the modes $F_{2g}^{(5)}$, $A_{g}^{(3)}$ and $E_{g}^{(4)}$ for which the accuracy is about 1~cm$^{-1}$.}
\begin{ruledtabular}
\begin{tabular}{ccccccccccc}
 &\multicolumn{5}{c}{10~K}  & \multicolumn{5}{c}{295~K} \\ \cmidrule(lr){2-6}  \cmidrule(lr){7-11} \\
 Mode & In$_2$O$_3$ bulk & $x = 0$ & $x = 0.05$
& $x = 0.08$ & $x = 0.1$ & In$_2$O$_3$ bulk & $x = 0$ & $x = 0.05$
& $x = 0.08$ & $x = 0.1$\\ \hline
 $F_{2g}^{(2)}$ & 114.2 & 115.2 & 115.4 & 116.7 & - & 112.7 & 112.3 & 112.9 & 113.0 & - \\
 $A_{g}^{(1)}$ & 138.0 & 138.7 & 136.8 & 136.5 & 136.5 & 135.7 & 135.0 & 134.0 & 133.7 & 133.8 \\
 $E_{g}^{(1)}$ & 175.8 & 176.4 & 178.4 & 180.4 & 181.6 & 173.2 & 172.5 & 175.0 & 176.8 & 178.4 \\
 $F_{2g}^{(4)}$ & 213.0 & 213.1 & 214.4 & 215.0 & 215.5 & 210.3 & 209.4 & 211.5 & 213.0 & 214.1 \\
 $F_{2g}^{(5)}$ & 219 & 220 & 220 & 221 & 222 & 216 & 216 & 216 & 218 & 220 \\
 $F_{2g}^{(8)}$ & 372.8 & 373.0 & 374.6 & 375.7 & 377.7 & 369.2 & 368.1 & 370.4 & 372.2 & 373.3 \\
 $F_{2g}^{(9)}$ & 396.2 & 397.0 & 400.0 & 402.2 & 405.6 & 392.6 & 391.3 & 395.0 & 398.6 & 400.2 \\
 $F_{2g}^{(10)}$ & 458.3 & 458.7 & 462.4 & 465.4 & 467.8 & 455.6 & 453.8 & 459.7 & 461.9 & 464.2 \\
 $F_{2g}^{(11)}$ & 472.7 & 473.3 & 476.4 & 479.8 & 483.2 & 469.0 & 467.8 & 472.5 & 475.1 & 476.9 \\
 $A_{g}^{(3)}$ & 502 & 500 & 505 & 508 & 512 & 495 & 496 & - & - & - \\
 $E_{g}^{(4)}$ & 599 & 602 & 605 & 611 & 616 & 599 & 599 & 600 & 601 & 602 \\
 $F_{2g}^{(14)}$ & 636.2 & 636.5 & 641.3 & 645.3 & 649.1 & 631.6 & 630.2 & 635.7 & 640.5 & 643.8 \\

\end{tabular}
\end{ruledtabular}
\end{table*}

Room-temperature (295~K) Raman spectra of bulk In$_2$O$_3$ and the nominally undoped (In,Ga)$_2$O$_3$ films are shown in Fig.~\ref{Spectra} for excitation close to the optical band gap of In$_2$O$_3$ (equivalent spectra acquired at low-temperature (10~K) can be found in the supplemental material). The In$_2$O$_3$ spectrum exhibits a series of phonon lines between 135 and 635~cm$^{-1}$ in accordance with the results of Ref.~\onlinecite{kranert_2014a} for excitation at 3.81~eV. The same optical phonon lines are resolved in the spectra of the (In$_{1-x}$Ga$_x$)$_2$O$_3$ samples with enlarged linewidth and with a shift to higher frequencies, with the exception for a few low-intensity and high-frequency modes. This finding, together with the structural analysis from Ref.~\onlinecite{papadogianni_2021}, confirms the bixbyite structure of the alloy films for Ga contents up to 10\%. Note that some phonon lines could not be resolved in the spectra of the alloy samples due to their broadening in combination with the relative small scattering volume in the thin films. Furthermore, no splitting could be observed for the phonon modes from the alloy samples, indicating a one-mode behavior of the corresponding phonon modes (see, e.g., Refs.~\onlinecite{grille_2000} and \onlinecite{davydov_2002}). From the symmetry of the bixbyite In$_2$O$_3$ crystal structure (space group 206, $Ia\overline{3}$\citep{bierwagen_2015}), the expected Raman active phonon modes are:\citep{white_1972}
 
\begin{equation}
\mathit{\Gamma} = 4A_g + 4E_g + 14F_{2g}.
\label{eq:symmetry}
\end{equation} 

Combining Raman spectra investigated at excitation wavelengths of 325 and 473~nm, all of the expected 22 Raman active modes\citep{garcia-domene_2012} can be observed for binary In$_2$O$_3$, assuming that the Raman line at 310~cm$^{-3}$ actually consists of a superposition including four overlapping modes. Symmetry assignments for most of the modes were done by the analysis of polarized Raman spectra from bulk In$_2$O$_3$, described in more detail in the supplemental material.
A reliable evaluation of the dependence on the Ga content has been possible for the $A_g^{(1)}$, $E_g^{(1)}$, $F_{2g}^{(8)}$, $F_{2g}^{(9)}$, and $F_{2g}^{(14)}$ modes.\citep{garcia-domene_2012} The shifts of their Raman peak frequencies with respect to the corresponding In$_2$O$_3$ frequencies are shown in Fig.~\ref{Phonons} as a function of the Ga content. The modes $E_g^{(1)}$, $F_{2g}^{(8)}$, $F_{2g}^{(9)}$, and $F_{2g}^{(14)}$ exhibit a nearly linear dependence on the Ga content. Only for the $A_g^{(1)}$ mode, in contrast, a redshift instead of a blueshift is found with a very weak dependence on the Ga content. As a consequence, the frequency shifts of all other phonon modes shown in Fig.~\ref{Phonons} are, thus, identified as being suitable for the determination of the Ga content in cubic (In,Ga)$_2$O$_3$ by Raman spectroscopy. In this respect, the $F_{2g}^{(14)}$ mode at 635~cm$^{-1}$ is best suited, i.e., has the strongest absolute dependence on the Ga content (see Fig.~\ref{Phonons}). The relative blueshift normalized to the corresponding phonon frequency in In$_2$O$_3$ however is slightly larger for some modes at lower frequencies, since those might be more sensitive to the occupation of the group III lattice sites, according to the relatively large masses of In and Ga. A list of low- and room-temperature phonon frequencies containing all modes which could be analyzed in an unambiguous manner for at least one alloy film is presented in Table~\ref{PhononFrequencies}. The nearly linear dependencies observed for most phonon frequencies (see Fig.~\ref{Phonons}) provide evidence for the absence of a solubility limit regarding the substitutional incorporation of Ga on group III lattice sites in In$_2$O$_3$ for Ga contents up to 10\%. 
In Ref.~\onlinecite{regoutz_2015}, in contrast, a blueshift for one particular phonon peak (actually consisting of a superposition of A$_g$ and  F$_g$ phonon modes\citep{kranert_2014a}) has been observed only up to Ga contents of 6\% for sintered ceramic alloy samples. For larger nominal Ga contents, the constant phonon frequency could be explained by phase separation as confirmed by X-ray diffraction measurements. Consequently, our results provide evidence for a higher Ga solubility limit for alloy films grown by MBE under the presented conditions.
The phonon frequencies are expected to depend on the residual lattice strain in the alloy films. As mentioned above, a maximal tensile in-plane strain level of about 0.2\% has been found for In$_2$O$_3$ films grown under the same conditions. For such a strain level, a redshift of the phonon modes by roughly 0.35\% has been observed for binary In$_2$O$_3$ films (not shown here). Assuming the same relative strain induced frequency shift for the alloy films, the data shown in Fig.~\ref{Phonons} are displayed with error bars, which indicate the correction for the expected strain-induced redshift on the high-frequency side. A confirmation of this assumption can be seen in Fig.~\ref{Phonons}, where the strain corrected mode frequencies for the In$_2$O$_3$ film approaches the values of the bulk In$_2$O$_3$ sample. In contrast to the optical absorption onset, the influence of free carriers on the phonon modes in centrosymmetric bixbyite films can be neglected for our analysis.\citep{ramsteiner_2020a}

In conclusion, the onset of optical absorption in cubic (In,Ga)$_2$O$_3$ alloy films depends linearly on the Ga content ($x$) with a blueshift of up to 150~meV for $x = 0.1$. Consistently, the fundamental band gap exhibits a blueshift with increasing Ga content. Similar to the case of binary In$_2$O$_3$, the absorption onset and the effective electron mass depend on the doping level. The optical phonon modes of cubic (In,Ga)$_2$O$_3$ exhibit individual dependencies on the Ga content with the absence of a detectable mode splitting indicating a one-mode alloy behavior. Several phonon modes appear to be suitable for the determination of the Ga content in (In,Ga)$_2$O$_3$ films by Raman spectroscopy.

See supplementary material for a more detailed description of the fitting procedure of the HAXPES data. Polarized Raman spectra for phonon assignment and low-temperature Raman spectra are included as well. In addition, point-by-point fits to raw ellipsometric data for the infrared as well as visible \& UV spectral range can be found there.

We thank Dr. Paulo Santos for a critical reading of the manuscript and Dr. Zbigniew Galazka for providing us with the state-of-the-art In$_2$O$_3$ bulk reference sample. This work was performed in the framework of GraFOx, a Leibniz-ScienceCampus partially funded by the Leibniz association. J.\ F.\ gratefully acknowledges the financial support by the Leibniz Association. We are grateful to HiSOR, Hiroshima University, and JAEA/SPring-8 for the development of HAXPES at BL15XU of SPring-8. The HAXPES measurements were performed under the approval of the NIMS Synchrotron X-ray Station (Proposal No. 2018A4601, 2018B4600 and 2019B4602).

The data that support the findings of this study are available from the corresponding author upon reasonable request.

%======================================================================

%======================================================================

%\bibliography{IGO_fundamental}


\end{document}

\title{Supplemental Material: Band gap widening and behavior of Raman-active phonon modes of cubic single-crystalline (In,Ga)$_2$O$_3$ alloy films}

\author{Johannes Feldl}
%\thanks{These two authors contributed equally}
\affiliation{Paul-Drude-Institut f\"ur Festk\"orperelektronik, Leibniz-Institut im Forschungsverbund Berlin e.~V., Hausvogteiplatz 5--7, 10117 Berlin, Germany}
\author{Martin Feneberg}
%\thanks{These two authors contributed equally}
\affiliation{Institut f\"ur Physik, Otto-von-Guericke-Universit\"at Magdeburg, Universit\"atsplatz 2, 39106 Magdeburg, Germany}
\author{Alexandra Papadogianni}
%\author{Shigenori Ueda}
%\affiliation{National Institute for Materials Science, Tukuba, Ibaraki 305-0044, Japan}
\author{Jonas L\"ahnemann}
\affiliation{Paul-Drude-Institut f\"ur Festk\"orperelektronik, Leibniz-Institut im Forschungsverbund Berlin e.~V., Hausvogteiplatz 5--7, 10117 Berlin, Germany}
\author{Takahiro Nagata}
\affiliation{National Institute for Materials Science, Tsukuba, Ibaraki 305-0044, Japan}
\author{Oliver Bierwagen}
\affiliation{Paul-Drude-Institut f\"ur Festk\"orperelektronik, Leibniz-Institut im Forschungsverbund Berlin e.~V., Hausvogteiplatz 5--7, 10117 Berlin, Germany}
\author{R\"udiger Goldhahn}
\affiliation{Institut f\"ur Physik, Otto-von-Guericke-Universit\"at Magdeburg, Universit\"atsplatz 2, 39106 Magdeburg, Germany}
\author{Manfred Ramsteiner}
\email{ramsteiner@pdi-berlin.de}
\affiliation{Paul-Drude-Institut f\"ur Festk\"orperelektronik, Leibniz-Institut im Forschungsverbund Berlin e.~V., Hausvogteiplatz 5--7, 10117 Berlin, Germany}

	\begin{abstract}
		{}
	\end{abstract}
	
	\maketitle
	\onecolumngrid

%\subsection{Computational Details}

For our investigation of the electronic state of the (In,Ga)$_2$O$_3$ alloy we employed hard X-ray photoelectron spectroscopy (HAXPES) with $h\nu = 5.95$~keV. The corresponding inelastic mean free path (IMFP: $\lambda$) of HAXPES for the In 3d core-level photoemission calculated by the Tanuma-Powell-Penn-2M\citep{tanuma_2006} is $\lambda = 7.29$~nm. Therefore, our HAXPES measurements probed approximately 21~nm (probing depth: $3\times$IMFP\citep{powell_1999,tanuma_2006}) below the sample surface, which can reduce the effect of the surface Fermi level pinning of In$_2$O$_3$.

\begin{figure}[b]%{r}{0.55\textwidth}
%\vspace*{-0.5cm}
\includegraphics*[width=0.65\textwidth]{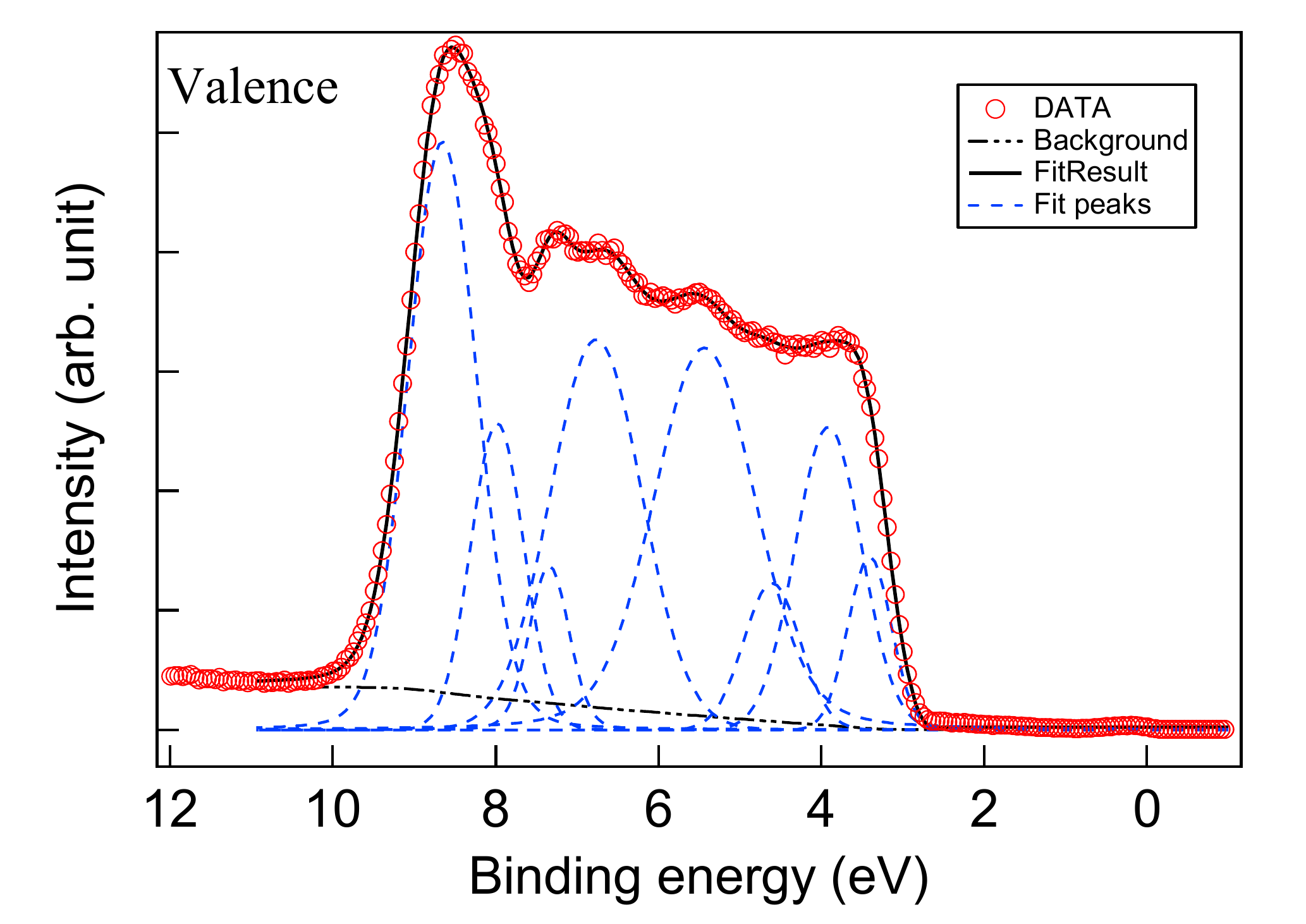}
\caption{HAXPES valence band data and corresponding Voigt multiple-peak fitting functions. The Fermi energy corresponds to an energy of 0~eV.} 
\label{sup_valence}
%\vspace*{-0.5cm}
\end{figure}

In this paper, we use the multi-peak fitting for the valence band region, as shown in Fig.~\ref{sup_valence}. By using the base peaks based on the Voigt function, we first fitted the valence band structure. Then, by considering the in-gap state and energy resolution ($\text{wid}$) using the Fermi edge fitting function

\begin{equation}
\text{FE}(x) = \frac{1}{1 + \exp\left(4\frac{x-\text{pos}}{\text{wid}} \right)},
\label{eq:one}
\end{equation} 

the valence band maximum value ($\text{pos}$) was estimated. Here, $x$ denotes the binding energy and the total fit is described by: 

\begin{equation}
\text{FitResult}(x) = \text{SumOfPeaksAndBackgrounds}(x)\times \text{FE}(x) + \text{FE\_offset},
\label{eq:two}
\end{equation} 

with intensity offset $\text{FE\_offset}$.

\begin{figure}[h]%{r}{0.55\textwidth}
%\vspace*{-0.5cm}
\includegraphics*[width=0.65\textwidth]{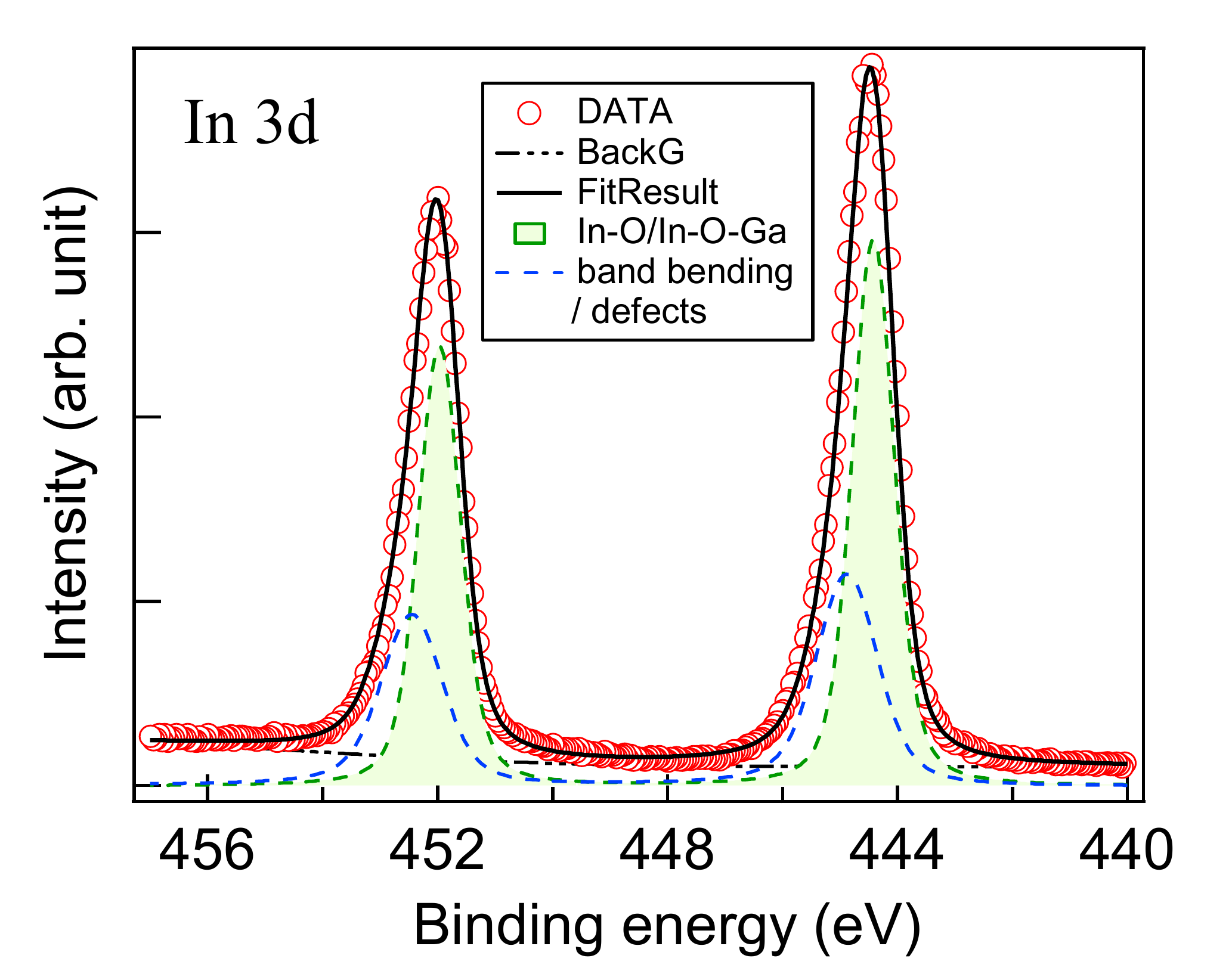}
\caption{In 3d core-level spectrum with corresponding two-peaks fitting functions. The binding energy is given relative to the Fermi level.} 
\label{sup_In3d}
%\vspace*{-0.5cm}
\end{figure}

The In 3d core-level spectra consist of In-O/In-O-Ga bonds and the effects of band bending corresponding to the Fermi level pinning and defective InOx bonds. To simplify these bonding states, we used two peaks for fitting of the In 3d core-level spectrum, as shown in Fig.~\ref{sup_In3d}.

%\subsection{Geometry}

%======================================================================
 %%supplemental Fig. 1
%%======================================================================

\begin{figure}[h]%{r}{0.55\textwidth}
%\vspace*{-0.5cm}
\includegraphics*[width=0.65\textwidth]{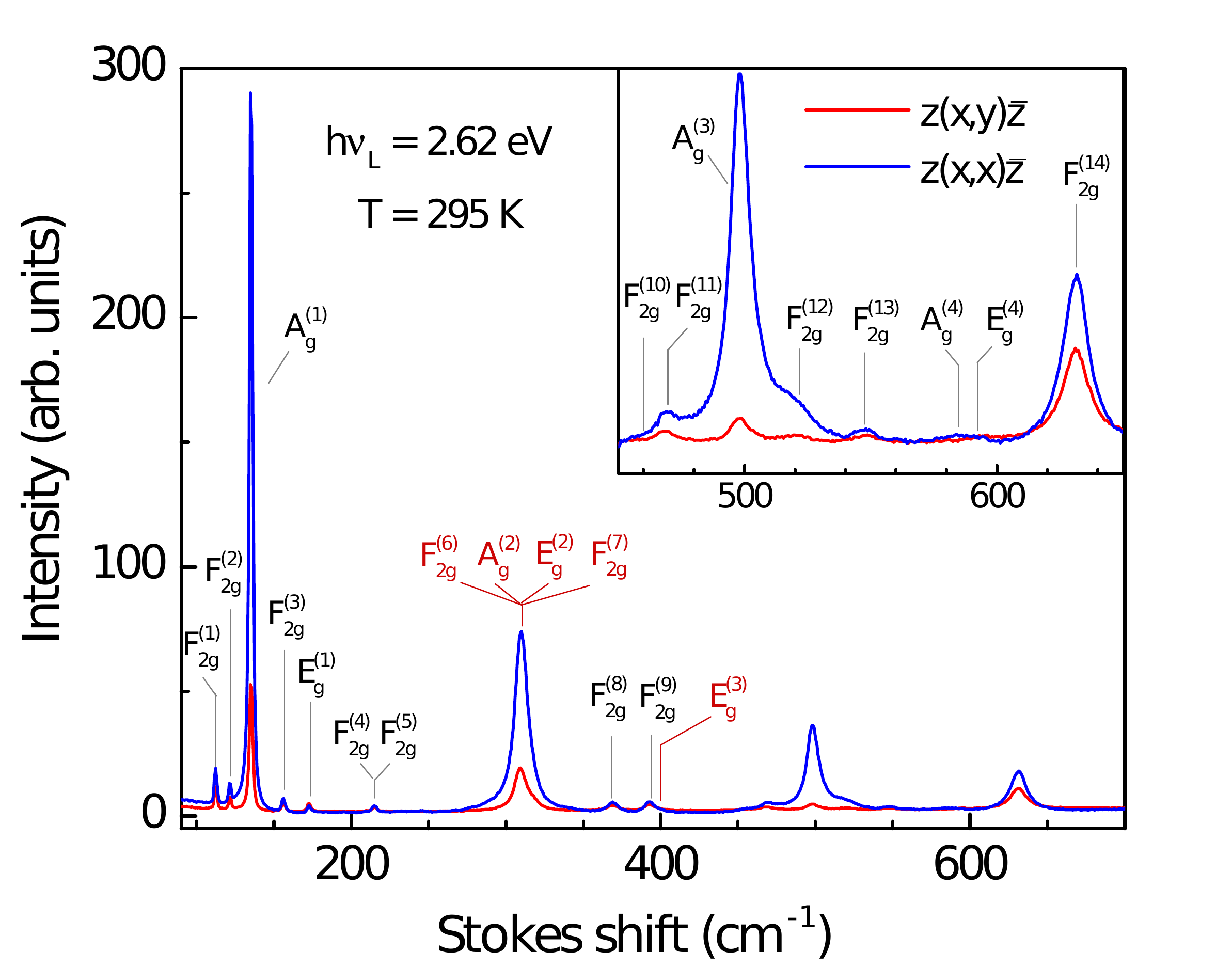}
\caption{Polarization dependent Raman measurements for the symmetry determination of the phonon modes. Modes which could not be identified unambiguously or are not present under the given experimental conditions are marked with red labels. The inset shows a magnification of the high frequency spectra.} 
\label{polarization}
%\vspace*{-0.5cm}
\end{figure}

For the assignment of the observed phonon lines, polarized Raman spectra were acquired at room-temperature for In$_2$O$_3$ bulk crystals with an excitation wavelength of 473~nm. While a parallel configuration allows for all modes ($A_g$, $E_g$ and $F_{2g}$) to be detected, a crossed configuration strongly suppresses $A_g$ modes while leaving $E_g$ ones mostly not affected.\citep{kranert_2014b} The remaining contribution of the $A_g$ modes most likely results from an imperfect alignment of the waveplates as well as a mixed output laser polarization.  Fig.~\ref{polarization} shows such polarization dependent measurements. Note that $z$ in the indicated scattering configurations corresponds to the [111] crystal orientation, whereas $x$ and $y$ are two orthogonal directions in the \{111\} plane.\citep{kranert_2014b} Not all expected modes are detectable under the presented experimental conditions or can be resolved unambiguously. These modes are displayed with a red label in Fig.~\ref{polarization}. The inset of Fig.~\ref{polarization} shows a magnification of the high frequency modes. Here, a very weak but reproducible peak in the binary In$_2$O$_3$ bulk sample appears at 585~cm$^{-1}$ for the $z(x,x)\overline{z}$ configuration. Hence, we assign this peak to the $A_g^{(4)}$ symmetry.

\begin{figure}%{r}{0.55\textwidth}
%\vspace*{-0.5cm}
\includegraphics*[width=0.6\textwidth]{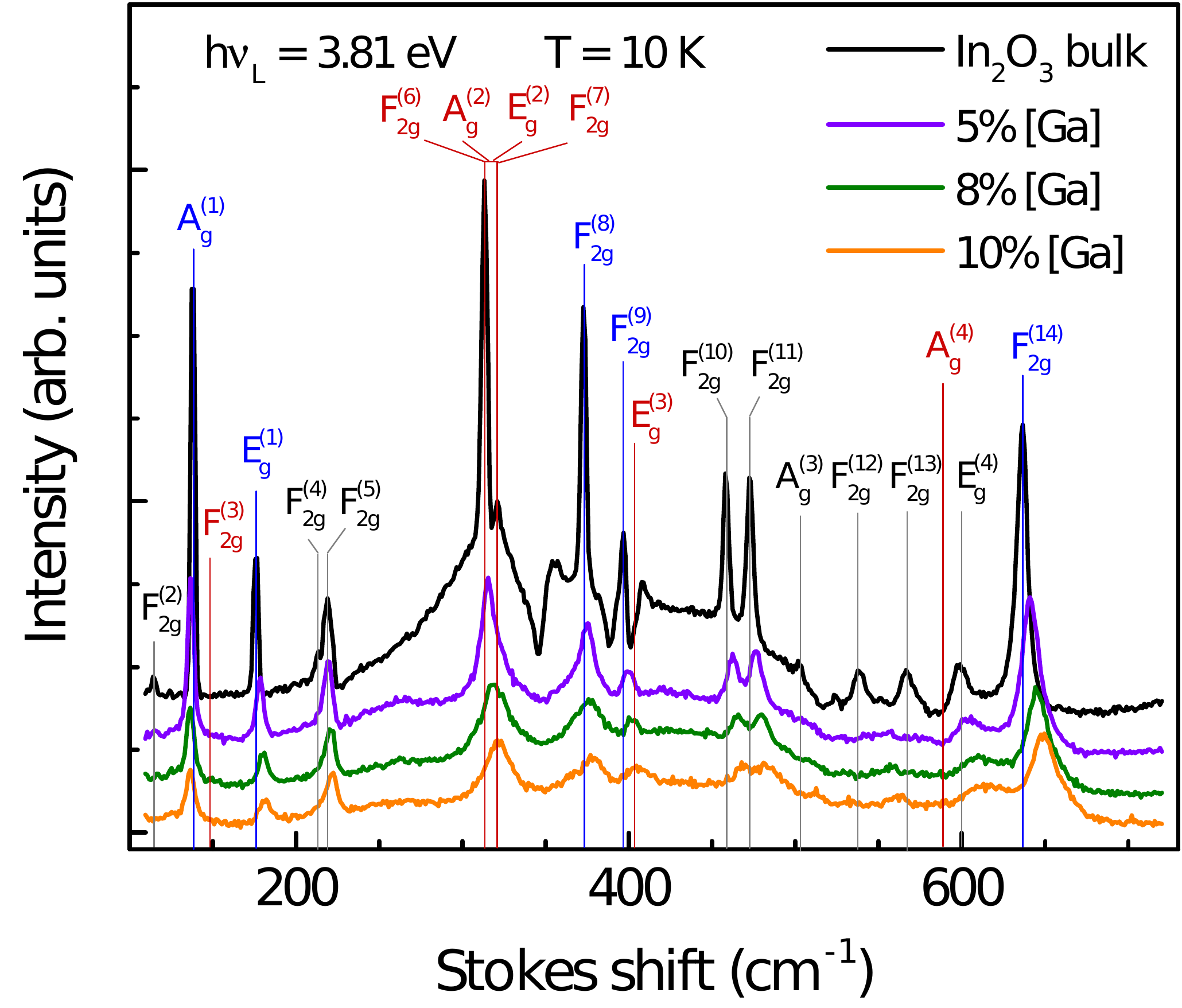}
\caption{Low-temperature (10~K) Raman spectra of (In$_{1-x}$Ga$_x$)$_2$O$_3$ films for different nominal Ga contents $x$ and a reference In$_{2}$O$_3$ bulk sample excited at a photon energy of 3.81~eV. The phonon modes marked by blue vertical lines and labels are the ones best suited for investigations on (In$_{1-x}$Ga$_x$)$_2$O$_3$ films at room-temperature. The remaining modes are marked by black color, except for those, not present in any of the samples under the given experimental conditions or whose symmetry identification was not unambiguous, marked in red.} 
\label{Spectra10K}
%\vspace*{-0.5cm}
\end{figure}

Complementary to the main text, Fig.~\ref{Spectra10K} shows low-temperature (10~K) Raman spectra which resemble the ones measured at room-temperature. However, the phonon lines are blueshifted and exhibit a smaller linewidth at 10~K.

%======================================================================
 %%supplemental Fig. 2
%%======================================================================

%======================================================================
 %%supplemental Fig. 3
%%======================================================================    

%======================================================================
 %%supplemental Fig. 4
% IGORamanSpectra from \InGa2O3\RT_325nm_Modes/July2020_allSamples_one_cali_RT/Figures
%%======================================================================

\begin{figure}[h]%{r}{0.55\textwidth}
%\vspace*{-0.5cm}
\includegraphics*[width=0.9\textwidth]{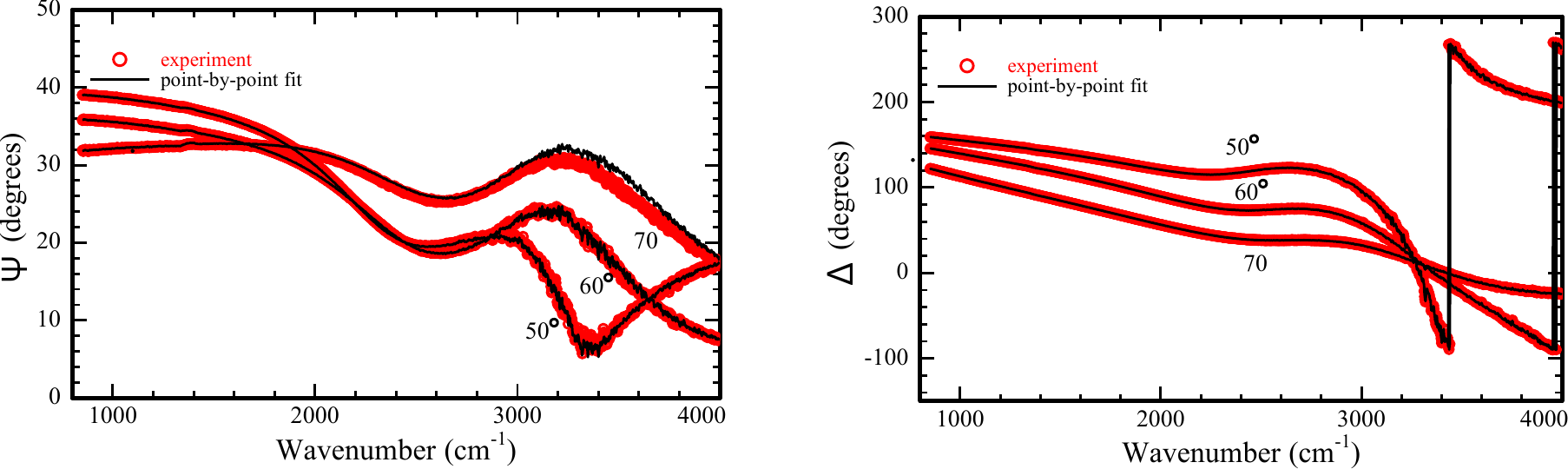}
\caption{Infrared point-by-point fit of raw ellipsometric data $\Psi$ and $\Delta$ of the as-grown (In,Ga)$_2$O$_3$:Sn with a carrier concentration of $4.4\cdot 10^{19}$~cm$^{-3}$.} 
\label{IRag}
%\vspace*{-0.5cm}
\end{figure}

\begin{figure}[h]%{r}{0.55\textwidth}
%\vspace*{-0.5cm}
\includegraphics*[width=0.9\textwidth]{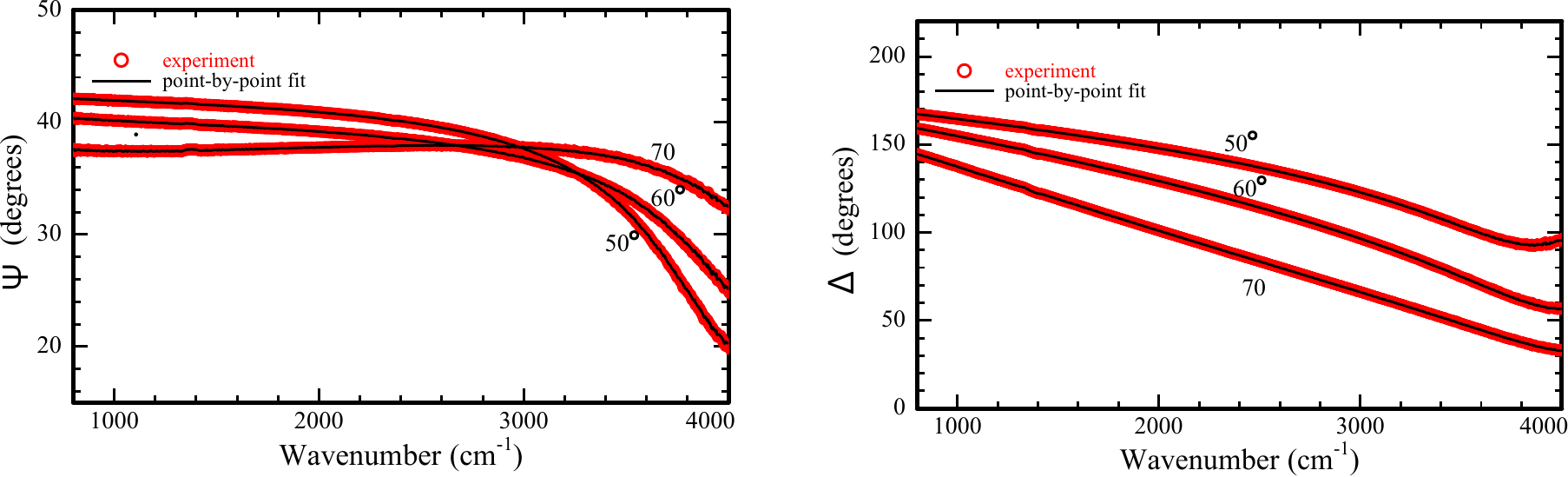}
\caption{Infrared point-by-point fit of raw ellipsometric data $\Psi$ and $\Delta$ of the oxygen-annealed (In,Ga)$_2$O$_3$:Sn with a carrier concentration of $16.5\cdot 10^{19}$~cm$^{-3}$.} 
\label{IRO2}
%\vspace*{-0.5cm}
\end{figure}

\begin{figure}[h]%{r}{0.55\textwidth}
%\vspace*{-0.5cm}
\includegraphics*[width=0.9\textwidth]{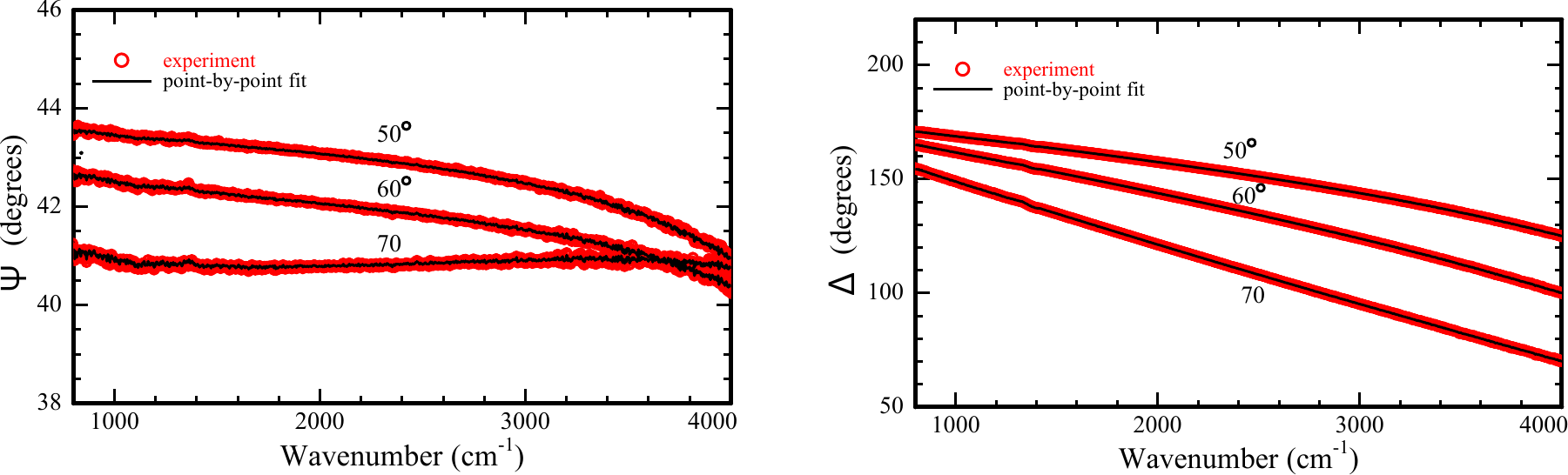}
\caption{Infrared point-by-point fit of raw ellipsometric data $\Psi$ and $\Delta$ of the vacuum-annealed (In,Ga)$_2$O$_3$:Sn with a carrier concentration of $35.0\cdot 10^{19}$~cm$^{-3}$.} 
\label{IRvac}
%\vspace*{-0.5cm}
\end{figure}

\begin{figure}[h]%{r}{0.55\textwidth}
%\vspace*{-0.5cm}
\includegraphics*[width=0.9\textwidth]{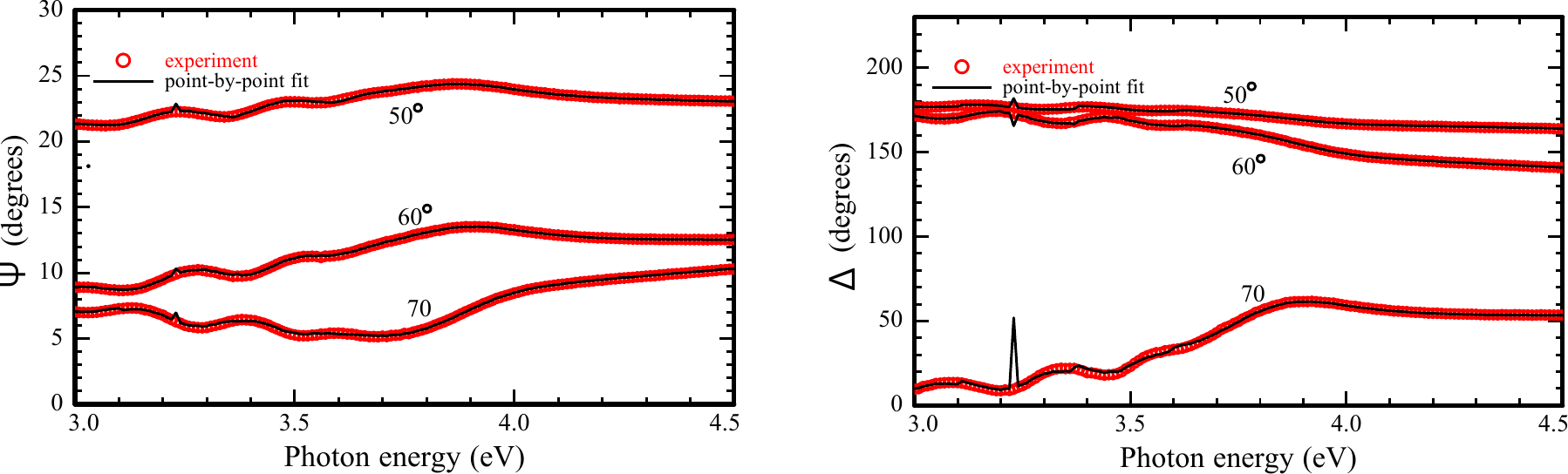}
\caption{Point-by-point fits of raw ellipsometric data $\Psi$ and $\Delta$ in the visible \& UV spectral range for the thin film with $x = 0.05$. A surface roughness layer of 3.0~nm was obtained.} 
\label{VIS_5}
%\vspace*{-0.5cm}
\end{figure}

\begin{figure}[h]%{r}{0.55\textwidth}
%\vspace*{-0.5cm}
\includegraphics*[width=0.9\textwidth]{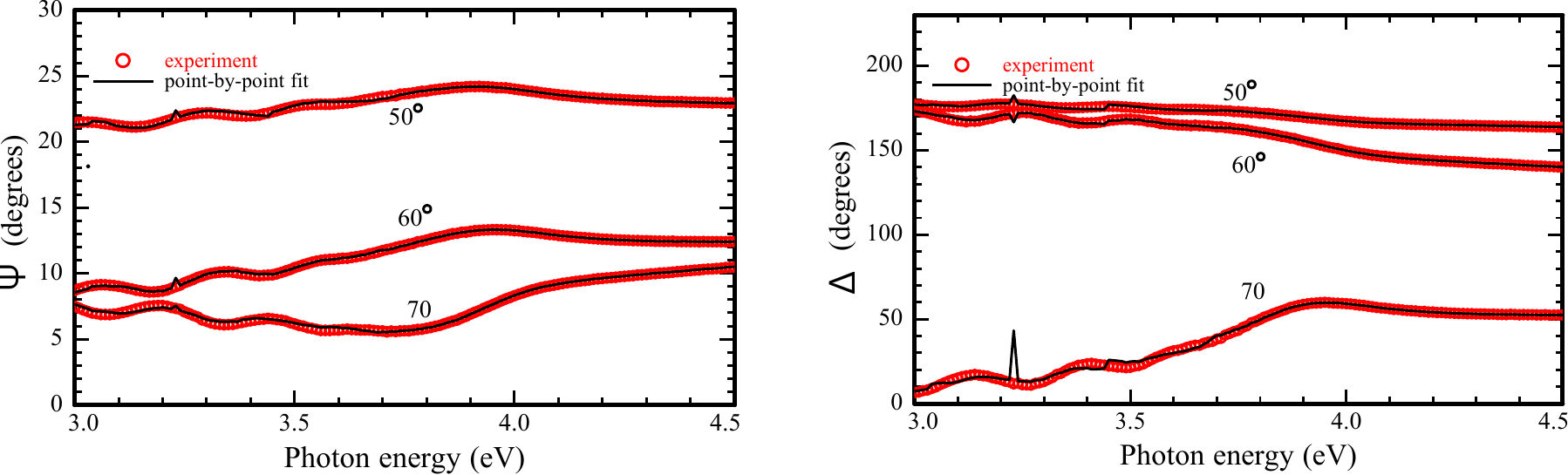}
\caption{Point-by-point fits of raw ellipsometric data $\Psi$ and $\Delta$ in the visible \& UV spectral range for the thin film with $x = 0.08$. A surface roughness layer of 3.3~nm was obtained.} 
\label{VIS_8}
%\vspace*{-0.5cm}
\end{figure}

\begin{figure}[h]%{r}{0.55\textwidth}
%\vspace*{-0.5cm}
\includegraphics*[width=0.9\textwidth]{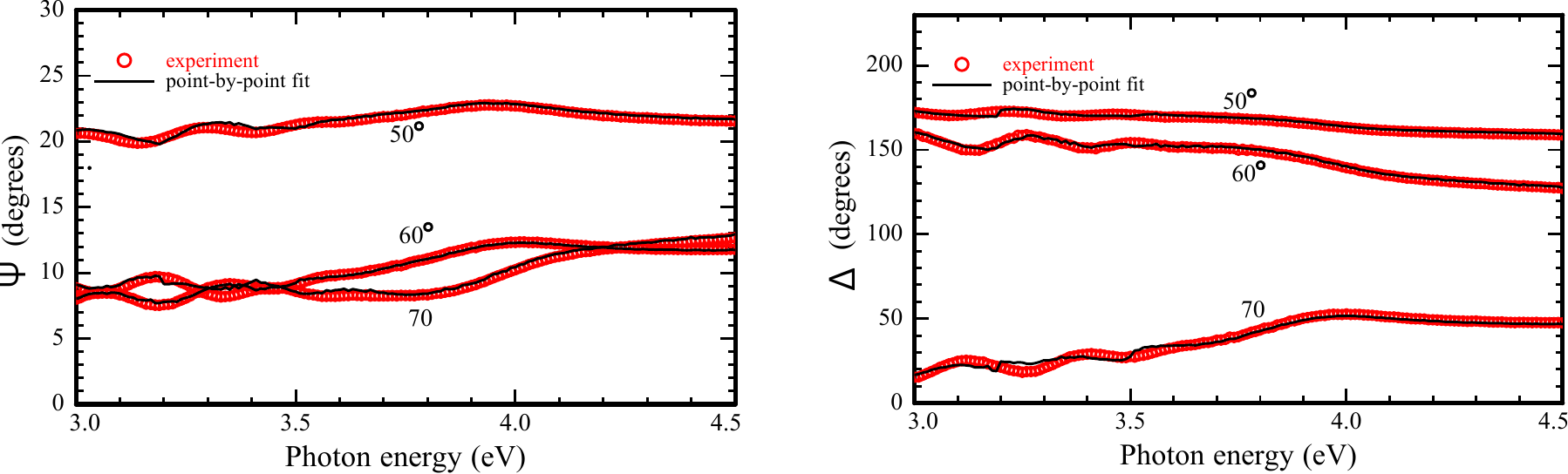}
\caption{Point-by-point fits of raw ellipsometric data $\Psi$ and $\Delta$ in the visible \& UV spectral range for the thin film with $x = 0.1$. A surface roughness layer of 7.5~nm was obtained.} 
\label{VIS_10}
%\vspace*{-0.5cm}
\end{figure}

In addition, we present the point-by-point fits to raw ellipsometric data $\Psi$ and $\Delta$. For the (In,Ga)$_2$O$_3$:Sn sample, Fig.~\ref{IRag}, Fig.~\ref{IRO2} and Fig.~\ref{IRvac} show the infrared data for the as-grown, oxygen-annealed and vacuum-annealed sample, respectively.
Corresponding data for the visible \& UV spectral range can be found in Fig.~\ref{VIS_5}, Fig.~\ref{VIS_8} and Fig.~\ref{VIS_10} for the thin films with Ga content $x$ of 0.05, 0.08 and 0.1 respectively. The surface roughness layer thicknesses obtained by model fitting agree well with the atomic force microscopy data presented by Papadogianni $\textit{e. al.}$\citep{papadogianni_2021}

%======================================================================
 %%supplemental Fig. 5
%%======================================================================   

%======================================================================
 %%supplemental Fig. 6
%%====================================================================== 

%======================================================================
 %%supplemental Fig. 7
%%====================================================================== 

%======================================================================
 %%supplemental Fig. 8
%%======================================================================  

%======================================================================
 %%supplemental Fig. 9
%%======================================================================  

%======================================================================
 %%supplemental Fig. 10
%%======================================================================  

%\bibliography{IGO_fun_sup}
